\documentclass[aps,prl,twocolumn]{revtex4-1}
\usepackage{amssymb}
\usepackage{amsmath,bm}
\usepackage{graphicx,color}
\usepackage{bbm}
\usepackage{multirow}
\newcommand{\B}[1]{{\bm{#1}}}
\newcommand{\sFrac}[2]{{\textstyle\frac{#1}{#2}}}

\begin{document}

\title{Effective Forces in Thermal Amorphous Solids with Generic Interactions}

\author{Giorgio Parisi}
\affiliation{Dipartimento di Fisica, Sapienza Universit\'a di Roma, INFN, Sezione di Roma I, IPFC -- CNR, Piazzale Aldo Moro 2, I-00185 Roma, Italy}

\author{Itamar Procaccia}
\affiliation{Department of Chemical Physics, the Weizmann Institute of Science, Rehovot 76100, Israel}

\author{Carmel Shor}
\affiliation{Department of Chemical Physics, the Weizmann Institute of Science, Rehovot 76100, Israel}

\author{Jacques Zylberg}
\affiliation{Department of Chemical Physics, the Weizmann Institute of Science, Rehovot 76100, Israel}

\begin{abstract}
In thermal glasses at temperatures sufficiently lower than the glass transition,
the constituent particles are trapped in their cages for sufficiently long time such that their {\em time-averaged positions} can be determined before diffusion and structural relaxation takes place. The effective forces are
those that hold these average positions in place. In numerical simulations the effective forces
$\B F_{ij}$ between any pair of particles can be measured as a time average of the {\em bare} forces $\B f_{ij}(\B r_{ij}(t))$. In general even if the bare forces come from two-body interactions, thermal dynamics
dress the effective forces to contain many-body interactions.  Here we develop the effective theory for systems with generic interactions, where the effective forces are derivable from an effective potential and in turn they give rise to an effective Hessian whose eigenvalues are all positive when the system is stable. In this Letter we offer analytic expressions for the effective theory, and demonstrate the usefulness and the predictive power of the approach.
\end{abstract}

\maketitle

{\bf Introduction}: In the last decade or two there has been great progress in understanding athermal amorphous solids at temperature $T = 0$ \cite{98ML,04ML,06TLB,09LP,10KLLP,11HKLP}. This progress was facilitated by the fact that particles’ positions $\{\B r_i\}_{i=1}^N$ are frozen at $T = 0$ and the knowledge of the microscopic (bare) forces is sufficient to develop a theory of the response of the materials to external mechanical or magnetic strains. The Hessian matrix, whose eigenvalues are semi-positive at $T=0$, supplies important information, leading to an athermal theory that provides good understanding of the density of states, of plastic events, and of the failure mechanisms of amorphous solids. Considerable progress was also achieved in understanding magnetic amorphous solids and cross effects between mechanical and magnetic responses \cite{13DHPS,15DHJMPS,16HPS}. These techniques fail however at finite temperature since the particle positions $\{\B r_i(t)\}_{i=1}^N$ fluctuate in time and inter-particle forces become dressed by dynamical effects. The Hessian matrix of a configuration at any given time $t$ contains negative eigenvalues and it cannot be
used to study stability and instabilities. The effective forces in thermal systems are determined by the momentum transferred when particles interact. Even when the bare forces are binary, the effective
forces contain ternary, quaternary and higher order terms \cite{16GLPPRR,18PPPRS}. In order to lift the methods that
were so useful at $T=0$ one needs a new idea: in thermal systems the particle positions are
indeed not stationary, but in glasses with large relaxation times one can determine the time-averaged positions before the onset of diffusion and much before the glass relaxes to thermodynamic equilibrium \cite{09MG}. The time averaged positions are trivially stationary in time, and we refer to such states as
``thermal mechanical equilibria" \cite{16DPSS}. In such states one can determine the renormalized force-laws that hold these average positions stable. These
renormalized force-laws will define an effective Hamiltonian and an effective Hessian matrix, offering a totally novel way to explore the stability, the responses to external strain and stresses, and the failure mechanisms of glassy materials at finite temperatures.

{\bf General Theory}: To develop the general effective theory consider a generic glass former composed
on $N$ particles in a volume $V$. The system is endowed with a bare potential $U(\{\B r_i(t)\})$.
It is customary to assume that the potential is a sum of binary interaction terms $\phi\left(\B r_{ij}(t)\right)$,
\begin{equation}
U(\{\B r_i(t)\}) = \sum_{<ij>} \phi\left(\B r_{ij}(t)\right)\ , \quad \B r_{ij} \equiv \B r_j-\B r_i \ ,
\end{equation}
where the symbol $<ij>$ denotes summation over interacting pairs only.
In this paper we consider bare potentials whose range exceeds the average inter-particle distance.
Lennard-Jones interactions are an example, but hard-spheres or even soft spheres are excluded, for reasons that will clarify soon. Such longer range interactions are referred to as ``generic".
The total bare force on a single particle and the bare inter-particle forces are
\begin{equation}
f^\alpha_i (\{\B r_k(t)\})\equiv -\frac{\partial U(\{\B r_k(t)\})}{\partial r_i^\alpha} \ , \quad f_{ij}^\alpha \left(\B r_{ij}(t)\right)= -\frac{\partial
\phi\left(\B r_{ij}\right)}{\partial r_{ij}^\alpha} \ .
\end{equation}
Note that Greek superscripts will be reserved below for Cartesian components of vectors and matrices. Lastly one can also define the bare Hessian matrix as
\begin{equation}
H_{ij}^{\alpha\beta} \equiv \frac{\partial f^\alpha_i}{\partial r_j^\beta} \equiv
-\frac{\partial^2 U}{\partial r_i^\alpha\partial r_j^\beta} \ .
\end{equation}
As said above, this Hessian matrix differs from its athermal counterpart in having negative eigenvalues.
The effective theory will cure this disadvantage.

When the amorphous solid under study allows the calculation of the average positions of the particles
within an interval of time $[0,\tau]$ such that each particle is only fluctuating within its own cage, we define the following time-stationary averages: The mean position of the $i$th particle $\B R_i$
is defined as
\begin{equation}
\B R_i \equiv \frac{1}{\tau} \int_0^\tau dt~ \B r_i(t) \ , \quad \B R_{ij} \equiv \B R_i -\B R_j \ .
\end{equation}
The mean force on the particle $i$ is defined as
\begin{equation}
\B F_i \equiv \frac{1}{\tau} \int_0^\tau dt~ \B f_i(\{\B r_k(t)\}_{i=1}^N) \ .
\label{meanF}
\end{equation}
The mean force between particles $i$ and $j$ is defined as
\begin{equation}
\B F_{ij} \equiv \frac{1}{\tau} \int_0^\tau dt~ \B f_{ij}\left(\B r_{ij}(t)\right) \ .
\label{meanFij}
\end{equation}
We note that
\begin{equation}
\B F_i=\sum_j \B F_{ij} =0 \quad \text{in thermal mechanical equilibrium.}
\label{equi}
\end{equation}
More importantly and less trivially, we stress that although $\B f_{ij}$ is only a function of $\B r_{ij}$, the effective force $\B F_{ij}$ is not binary, and in principle it can contain many-body interactions.

{\bf Theory for generic potentials}. For generic potentials we will write
\begin{equation}
\B r_{ij}(t)=\B R_{ij} + \B u_{ij}(t) \ ,
\label{exp}
\end{equation}
 and expand the wanted objects to any desired order in $u_{ij}$. We will examine the efficacy of this approach with
Lennard-Jones glasses below. Thus for example we will use Eq.~(\ref{exp}) in Eq.~(\ref{meanFij}). Denoting objects expanded to a desired order in $u_{ij}$ with a hat, we find
\begin{equation}
\hat F^\eta_{ij} = f_{ij}^\eta (\B R_{ij})
+\frac{1}{2} \frac{\partial^2f_{ij}^\eta}{\partial r_{ij}^\alpha\partial r_{ij}^\beta}\Big |_{\B R_{ij}}     \langle u_{ij}^\alpha(t) u_{ij}^\beta(t)\rangle +\cdots\ . \label{hatFij}
\end{equation}
where repeated indices are summed upon, angular brackets denote time average and ``$\cdots$" represent terms of higher order in $u_{ij}$ if such terms are deemed necessary.
We note that the linear term in $u_{ij}$ vanishes upon time averaging.  The derivatives are evaluated at the mean vector distance. It becomes clear now why this approach is not applicable to hard or soft spheres; typically the separations $R_{ij}$ exceed the range of interaction and the derivatives employed in Eq.~(\ref{hatFij}) do not exist. Similarly, affecting the same approximation to the definition of $\phi_{ij}$ we find that
\begin{equation}
\hat \Phi_{ij} = \phi(R_{ij}) + \frac{1}{2}\frac{\partial^2\phi}{\partial r_{ij}^\alpha\partial r_{ij}^\beta}\Big|_{\B R_{ij}} \langle u_{ij}^\alpha(t) u_{ij}^\beta(t)\rangle +\cdots\ ,
\label{meanphiij}
\end{equation}
 We note that the force between particles $i$ and $j$ in the present approximation
is {\bf not} a function of $R_{ij}$ only, since the cage fluctuations $\langle u_{ij}^\alpha u_{ij}^\beta\rangle$ depend on all the particles. This is where the many-body interactions are implicitly affecting the present approximation. Finally we note that the effective
inter-particle force is derivable from the effective potential,
\begin{equation}
\hat F^\eta_{ij} = -\frac{\partial \hat \Phi_{ij}}{\partial r_{ij}^\eta}\Big|_{\B R_{ij}} \ ,
\label{Ffromphi}
\end{equation}
if we adopt the convention that the cage fluctuations are taken as input numbers in Eq.~(\ref{meanphiij}) and are not derived. Similarly, with the same rule, the effective
Hessian is obtained by a second derivative of Eq.~(\ref{meanphiij}) or as a first
derivative of Eq.~(\ref{Ffromphi}). Below we ascertain that the effective forces compute this
way sum up to zero, $\sum_j \hat{\B F}_{ij}=0$ and that the effective Hessian has no negative
eigenvalues, as required.

{\bf Testing in Lennard-Jones models}: Before proceeding we should test the quality of the
truncated expansion in standard models of glass formers.
For the numerical experiments we employ a generic glass former in 2-dimensions in the form of the
Kob-Andersen binary mixture\cite{95KA,09BSPK} of Lennard-Jones bare interactions cut off at $r_{\rm co}$ with four smooth derivatives. The potentials have the analytic form
\begin{equation}\label{potential}
\phi\!\left(\!\sFrac{r_{ij}}{\lambda_{ij}}\!\right)\! =\!
\left\{\begin{array}{l}
\!\varepsilon\!\left[\left(\!\frac{\lambda_{ij}}{ r_{ij}}\!\right)^{12}\!-\left(\!\frac{\lambda_{ij}}{ r_{ij}}\!\right)^{6}
+ \displaystyle{\sum_{\ell=0}^{6}}c_{\ell}\left(\!\sFrac{r_{ij}}{\lambda_{ij}}\!\right)^{\ell}\right]
  \!, \hskip 0.0 cm\frac{r_{ij}}{\lambda_{ij}}<\frac{r_{\rm co}}{\lambda} \\
\quad\quad\quad\quad\quad 0
    \hskip 3.4 cm  , \frac{r_{ij}}{\lambda_{ij}} \ge\frac{r_{\rm co}}{\lambda}
\end{array}
\right.\!\!
\end{equation}
The unit of length $\lambda$ is set to be the interaction length scale of two small particles. We solve the coefficients $c_{\ell}$ for the potential such
that the potential has its minimum $\phi=-\varepsilon$ at $r_{\rm min}/\lambda_{ij}=2^{1/6}$ and it vanishes with four continuous derivatives at $r_{\rm co}/\lambda_{ij}=2.5$.
In these units the Boltzmann constant $k_B = 1$.

 The simulations are performed in an NVT ensemble with density $\rho=1.162$ using a modified Berendsen thermostat which couples a constant number of particles to the bath, regardless of the system size \cite{10KLPZ}. In this method the temperature is defined by the average kinetic energy per particles, and a constant temperature is achieved by velocity re-scaling. Using samples with $N=500$ particles we first equilibrate a liquid at $T=0.5$ (in the usual LJ units)
than cool the system slowly at $\dot{T}=10^{-6}$ down to $T=10^{-6}$.
Lastly the samples are heated up at the same rate up to the desired temperatures $T$, where the averaged positions and moments of displacement are calculated.
\begin{figure}
\includegraphics[width=.45\textwidth]{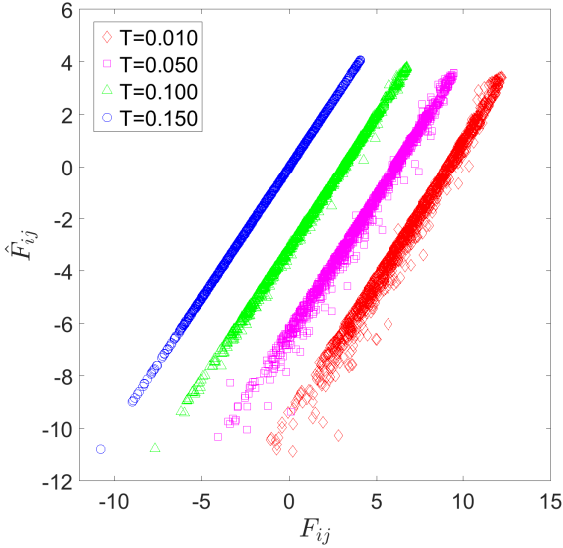}
\caption{Comparison of the effective inter-particle forces computed from Eq.~(\ref{hatFij}) with the ``exact" time averaged forces Eq.~(\ref{meanFij}). Here are shown all the non zero interactions
at four different temperatures. The plots were shifted for clarity.}
\label{compare}
\end{figure}
Using this model we determined the effective inter-particle forces $\B F_{ij}$ using the time
average Eq.~({\ref{meanFij}) and the approximation Eq.~(\ref{hatFij}). In Fig.~\ref{compare} we compare one against the other for four
different temperatures, $T=0.01$, $T=0.05$ and $T=0.1$ and $T=0.15$ in Lennard-Jones units.
Computing the Pearson correlation coefficients for these data we find the values $R^2 = 0.999,0.992
,0.979$ and 0.965 respectively. Obviously the error grows with the temperature indicating that at higher temperature one
will need higher corrections beyond the second order. Notwithstanding, we consider this results as an excellent support for the approach and proceed now to test its power in determining the mechanical
properties of the thermal amorphous solid.

{\bf The shear modulus}:
To demonstrate the usefulness of the approach we turn now to the computation of the shear-modulus.
Before going to the thermal case, we recall that at athermal conditions ($T=0$) the shear modulus has the exact representation:
\begin{equation}
\mu(T=0)\! = \!\frac{\partial \sigma_{xy}}{\partial \epsilon_{xy}}-\sum\limits_{i,\alpha}\!\!\sum\limits_{j,\beta}\Xi_i^\alpha (H_{ij}^{\alpha\beta})^{-1}
\Xi_j^\beta  \ , \quad \Xi_i^\alpha\equiv \frac{\partial \sigma_{xy}}{\partial r_i^\alpha} \ .\label{muath}
\end{equation}
where the first term is the Born approximation.  We note that here and below the Hessian matrix has zero eigenvalues due to Goldstone modes, and theses should be removed before inversion. In the thermal case
\cite{89Lutsko,barrat2006microscopic,12Yos} the Born term is corrected by stress fluctuations:
\begin{align}
\mu =\left< \frac{\partial \sigma_{xy}}{\partial \epsilon_{xy}} \right>  - \frac{V}{k_B T}\left( \langle \sigma_{xy}^2 \rangle - \langle \sigma_{xy} \rangle^2\right)
+ 2 \rho k_B T \ .
\label{eq:muDef}
\end{align}

As before we determine
an ``effective shear modulus" $\hat \mu$ by using the stationary positions
$\B R_i$ and the cage fluctuations $\B u_i$. We begin by expressing the stress $\sigma\equiv \sigma_{xy}$:
\begin{align}
&&\sigma (\{\B r_i(t)\})=\sigma(\{\B R_i\}) +\sum\limits_{i,\alpha}\frac{\partial \sigma}{\partial r_i^{\alpha}}\Big|_{\{\B R_i\}}u_i^{\alpha}\nonumber\\
&&+
\frac{1}{2}\sum\limits_{i,\alpha}\sum\limits_{j,\beta}\frac{\partial^2\sigma}{\partial r_i^{\alpha}\partial r_j^{\beta}}\Big|_{\{\B R_i\}}u_i^{\alpha}u_j^{\beta}
+\dots \: \: .
\end{align}

The first term in Eq.\eqref{eq:muDef} is the Born term, and using Eq.(\ref{meanphiij}) it can be expanded again in the cage
fluctuations:
\begin{eqnarray}
\hat \mu_{Born} =
&&\frac{1}{V}\frac{\partial^2 U }{\partial \epsilon_{xy}^2}\Big|_{\{\B R_i\}} \label{Borntay}\\
&&+\frac{1}{V}\frac{1}{2} \sum_{<ij>} \Big[   \frac{\partial^2 }{\partial r_{ij}^{\alpha}\partial r_{ij}^{\beta}} \frac{\partial^2 U }{\partial \epsilon_{xy}^2} \Big|_{\B R_{ij}} \langle u_{ij}^{\alpha}u_{ij}^{\beta} \rangle  \Big] +\ldots , \nonumber
\end{eqnarray}
The second term in Eq.\eqref{eq:muDef} contains the second moment $\langle \sigma^2_{xy} \rangle$. When we expand
this object we need to take into consideration the first order fluctuation  $u_i^\alpha$:
\begin{eqnarray}
&&\big[ \sigma(\{\B r_i(t)\}) \big]^2 =
\big[ \sigma(\{\B R_i\}) \big]^2
+ 2\sigma(\{\B R_i\})  \sum\limits_{i,\alpha}\frac{\partial \sigma}{\partial r_i^{\alpha}}\Big|_{\{\B R_i\}}u_i^{\alpha}\nonumber\\ &&
+ \frac{\sigma(\{\B R_i\}) }{2} \sum\limits_{i,\alpha}\sum\limits_{j,\beta}\frac{\partial^2\sigma}{\partial r_i^{\alpha}\partial r_j^{\beta}}\Big|_{\{\B R_i\}}u_i^{\alpha}u_j^{\beta}
\\ &&
+\sum\limits_{i,\alpha}\frac{\partial \sigma}{\partial r_i^{\alpha}}\Big|_{\{\B R_i\}}
\sum\limits_{j,\beta}\frac{\partial \sigma}{\partial r_j^{\beta}}\Big|_{\{\B R_i\}}u_i^{\alpha}u_i^{\beta}
+ \dots;
\nonumber
\end{eqnarray}
For the average of the second moment $\langle \sigma^2 \rangle$ the linear contribution in $u_i^\alpha$ vanishes, and we are left with:
\begin{eqnarray}
\label{eq:sigsqT}
&&\big< \big[ \sigma(\{\B r_i(t)\}) \big]^2 \big> =
\big[ \sigma(\{\B R_i\}) \big]^2 \nonumber \\ &&
+ \frac{\sigma(\{\B R_i\}) }{2}  \sum\limits_{i,\alpha}\sum\limits_{j,\beta}\frac{\partial^2\sigma}{\partial r_i^{\alpha}\partial r_j^{\beta}}\Big|_{\{\B R_i\}}
\left< u_i^{\alpha}u_j^{\beta} \right>
\\ &&
+\sum\limits_{i,\alpha}\frac{\partial \sigma}{\partial r_i^{\alpha}}\Big|_{\{\B R_i\}}
\sum\limits_{j,\beta}\frac{\partial \sigma}{\partial r_j^{\beta}}\Big|_{\{\B R_i\}}
\left< u_i^{\alpha}u_j^{\beta} \right>
+ \dots.
\nonumber
\end{eqnarray}
For $\langle \sigma \rangle^2$, the $\langle u_i^\alpha \rangle$ terms have already vanished before squaring the term, so we get:
\begin{eqnarray}
&&\big< \big[ \sigma(\{\B r_i(t)\} )  \big] \big>^2=
\big[ \sigma(\{\B R_i\}) \big]^2  \\ &&
+ \sigma(\{\B R_i\})   \sum\limits_{i,\alpha}\sum\limits_{j,\beta}\frac{\partial^2\sigma }{\partial r_i^{\alpha}\partial r_j^{\beta}}\Big|_{\{\B R_i\}}
\left< u_i^{\alpha}u_j^{\beta} \right> \nonumber
+ \dots.
\end{eqnarray}
Subtracting the last two equations we get the desired expression for the Taylor expansion of the fluctuation term in Eq.\eqref{eq:muDef}:
\begin{eqnarray}
\hat{\mu}_F
= \frac{V}{k_B T}
\sum\limits_{i,\alpha}\sum\limits_{j,\beta}\frac{\partial \sigma}{\partial r_i^{\alpha}}\Big|_{\{\B R_i\}} \left< u_i^{\alpha}u_j^{\beta} \right>
\frac{\partial \sigma}{\partial r_j^{\beta}}\Big|_{\{\B R_i\}}
 +\dots
\label{varSigTay}
\end{eqnarray}
Note that in this case the cage fluctuations are represented by the correlation matrix $C_{ij}^{\alpha \beta} = \langle u_i^\alpha u_j^\beta \rangle =  \langle (r^\alpha_i - R^\alpha_i)(r_j^\beta - R^\beta_j) \rangle$ , and not by the ``pair fluctuations" $\langle u_{ij}^\alpha u_{ij}^\beta \rangle$ as in Eq. \eqref{hatFij} and Eq.\eqref{meanphiij} above.
We expect this correlation to be proportional to $V^{-1}$ due to the central limit theorem,
cancelling the explicit volume factor on the RHS of Eq.~(\ref{varSigTay}).

Recalling that the correlation function $C_{ij}^{\alpha \beta}$ is intimately related to the inverse (bare) Hessian \cite{dauchot2012pedagogical}, i.e. that
\begin{equation}
\B C = k_BT \B H^{-1}\ ,
\end{equation}
we appreciate the apparent structural relationship between Eqs.~(\ref{varSigTay}) and (\ref{muath}).
Finally using Eq.\eqref{Borntay} and Eq.~\eqref{varSigTay} we can write:
\begin{equation}
\hat{\mu}= \hat{\mu}_{Born}-\hat{\mu}_F.
\label{eq:hatMu}
\end{equation}

A comparison between the results of computing the shear modulus via Eq.~(\ref{eq:hatMu}) and Eq.(\ref{eq:muDef}) now called for.
For each Temperature $T\in[0.001..0.100]$ some 100 samples were prepared, and quenched separately from $T=0.5$ to $T=10^{-6}$ with quench rate of $\dot{T} = 10^{-6}$. Next, each sample was heated up to the desired temperature;  the average positions $\B R_i$ cage fluctuation correlations were calculated by averaging over $5\times10^5$ MD steps;  All together, each point in Fig.\ref{fig:muTay} was averaged over  $50-100$ samples. The results of the comparison are shown in Fig \ref{fig:muTay}. The line
drawn is just a guide for the eye.

\begin{figure}[h]
	\begin{center}
		\includegraphics[width=.45\textwidth]{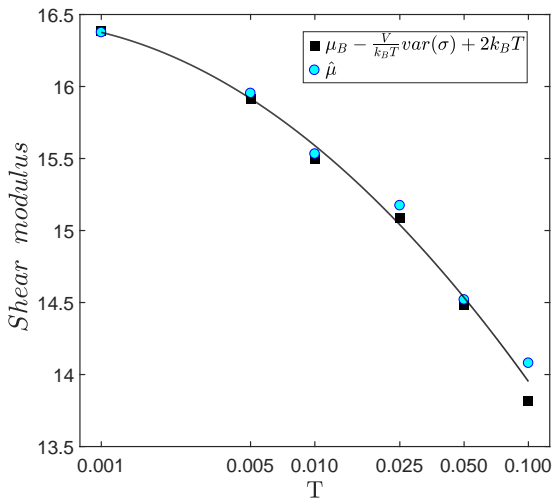} \\

	\end{center}
	\caption[$\mu$ and $\mu^{eff}$ for different temperatures]{Shear-modulus $\mu$ calculated from Eq.\eqref{eq:muDef} using stress-fluctuations (black squares), and from the effective theory using Eq.\eqref{eq:hatMu} (blue circles). The line is a guide for the eye.}
	\label{fig:muTay}
\end{figure}

{\bf Concluding remarks}:
The results obtained for the effective forces and the for shear modulus open up a new path
for studying the physics of amorphous solids at finite temperatures. After all, when we
discuss glasses at low temperatures, the average spatial structure is quite stable for a long time,
oblivious of the thermal agitation of particles within their cages. Do we really care about the
details of this thermal agitation?. The answer is yes and no. Yes, because this motion dresses
up the interactions between the particles, and the bare forces are no longer providing a
proper description of some important properties. Momentum transfer is taking place, and this
fact has consequences. On the other hand, the average positions of the particles within their
cages offers a skeleton for the theory of thermal amorphous solids in much the same way
as the frozen position at zero temperature. What we need to learn is how to take the pertinent
information into account for devising a good theory \cite{84SW,08RW}.

The present results indicate that at least in glasses where the average distance between particles
is within the range of the bare interactions, we can reach a theory by expanding objects around the average positions. In this paper we stopped at the first correction, limiting ourselves to low temperatures. We note that this is NOT a harmonic approximation of the bare potential; the theory presented above
calls for higher derivatives of the bare potential, up to fourth order already; for example the RHS of Eq.~(\ref{hatFij}) employs a third order derivative of the potential, and Eq.~(\ref{Borntay}) a fourth order. Higher order
truncations necessitated by higher temperatures will require more smooth derivatives in the bare
potential. We reiterate that the present approach will fail for hard or soft spheres, and also
for some inverse power law models where the exponent is too large. In such models the mean distance between particles will exceed
the range of interaction and we do not have the required derivatives of the bare
potential that are computed at the mean separations.

One important question remaining for future research is how to build up on the present
ideas a theory of instabilities and mechanical failure in thermal glasses. In athermal
conditions the eigenvalues of the Hessian matrix provided enormous insights on plastic
responses, density of states and shear banding instabilities. If we accept the view that
the random thermal motions within cages should be averaged over, then the effective
Hessian introduced above should be studied for the purpose of providing a similar
understanding in thermal systems. This and other related issues will be studied
in the near future.

\acknowledgments
We thank George Hentschel for some very helpful discussions at the initiation of this research. This work was supported
in part by the Israel Science Foundation (Israel Singapore Program), US-Israel Binational Science Foundation and the Joint Laboratory on ``Advanced and Innovative Materials" - Universita' di Roma ``La Sapienza" - WIS.

\bibliography{ALL}

\begin{thebibliography}{22}%
\makeatletter
\providecommand \@ifxundefined [1]{%
 \@ifx{#1\undefined}
}%
\providecommand \@ifnum [1]{%
 \ifnum #1\expandafter \@firstoftwo
 \else \expandafter \@secondoftwo
 \fi
}%
\providecommand \@ifx [1]{%
 \ifx #1\expandafter \@firstoftwo
 \else \expandafter \@secondoftwo
 \fi
}%
\providecommand \natexlab [1]{#1}%
\providecommand \enquote  [1]{``#1''}%
\providecommand \bibnamefont  [1]{#1}%
\providecommand \bibfnamefont [1]{#1}%
\providecommand \citenamefont [1]{#1}%
\providecommand \href@noop [0]{\@secondoftwo}%
\providecommand \href [0]{\begingroup \@sanitize@url \@href}%
\providecommand \@href[1]{\@@startlink{#1}\@@href}%
\providecommand \@@href[1]{\endgroup#1\@@endlink}%
\providecommand \@sanitize@url [0]{\catcode `\\12\catcode `\$12\catcode
  `\&12\catcode `\#12\catcode `\^12\catcode `\_12\catcode `\%12\relax}%
\providecommand \@@startlink[1]{}%
\providecommand \@@endlink[0]{}%
\providecommand \url  [0]{\begingroup\@sanitize@url \@url }%
\providecommand \@url [1]{\endgroup\@href {#1}{\urlprefix }}%
\providecommand \urlprefix  [0]{URL }%
\providecommand \Eprint [0]{\href }%
\providecommand \doibase [0]{http://dx.doi.org/}%
\providecommand \selectlanguage [0]{\@gobble}%
\providecommand \bibinfo  [0]{\@secondoftwo}%
\providecommand \bibfield  [0]{\@secondoftwo}%
\providecommand \translation [1]{[#1]}%
\providecommand \BibitemOpen [0]{}%
\providecommand \bibitemStop [0]{}%
\providecommand \bibitemNoStop [0]{.\EOS\space}%
\providecommand \EOS [0]{\spacefactor3000\relax}%
\providecommand \BibitemShut  [1]{\csname bibitem#1\endcsname}%
\let\auto@bib@innerbib\@empty
\bibitem [{\citenamefont {Malandro}\ and\ \citenamefont {Lacks}(1998)}]{98ML}%
  \BibitemOpen
  \bibfield  {author} {\bibinfo {author} {\bibfnamefont {D.~L.}\ \bibnamefont
  {Malandro}}\ and\ \bibinfo {author} {\bibfnamefont {D.~J.}\ \bibnamefont
  {Lacks}},\ }\href@noop {} {\bibfield  {journal} {\bibinfo  {journal} {Phys.
  Rev. Lett.}\ }\textbf {\bibinfo {volume} {81}},\ \bibinfo {pages} {5576}
  (\bibinfo {year} {1998})}\BibitemShut {NoStop}%
\bibitem [{\citenamefont {Maloney}\ and\ \citenamefont
  {Lema{\^{i}}tre}(2004)}]{04ML}%
  \BibitemOpen
  \bibfield  {author} {\bibinfo {author} {\bibfnamefont {C.}~\bibnamefont
  {Maloney}}\ and\ \bibinfo {author} {\bibfnamefont {A.}~\bibnamefont
  {Lema{\^{i}}tre}},\ }\href {\doibase 10.1103/PhysRevLett.93.195501}
  {\bibfield  {journal} {\bibinfo  {journal} {Physi. Rev. Lett.}\ }\textbf
  {\bibinfo {volume} {93}},\ \bibinfo {pages} {195501} (\bibinfo {year}
  {2004})}\BibitemShut {NoStop}%
\bibitem [{\citenamefont {Tanguy}\ \emph {et~al.}(2006)\citenamefont {Tanguy},
  \citenamefont {Leonforte},\ and\ \citenamefont {Barrat}}]{06TLB}%
  \BibitemOpen
  \bibfield  {author} {\bibinfo {author} {\bibfnamefont {A.}~\bibnamefont
  {Tanguy}}, \bibinfo {author} {\bibfnamefont {F.}~\bibnamefont {Leonforte}}, \
  and\ \bibinfo {author} {\bibfnamefont {J.~L.}\ \bibnamefont {Barrat}},\
  }\href {\doibase 10.1140/epje/i2006-10024-2} {\bibfield  {journal} {\bibinfo
  {journal} {Euro. Phys. J. E}\ }\textbf {\bibinfo {volume} {20}},\ \bibinfo
  {pages} {355} (\bibinfo {year} {2006})}\BibitemShut {NoStop}%
\bibitem [{\citenamefont {Lerner}\ and\ \citenamefont
  {Procaccia}(2009)}]{09LP}%
  \BibitemOpen
  \bibfield  {author} {\bibinfo {author} {\bibfnamefont {E.}~\bibnamefont
  {Lerner}}\ and\ \bibinfo {author} {\bibfnamefont {I.}~\bibnamefont
  {Procaccia}},\ }\href {\doibase 10.1103/PhysRevE.79.066109} {\bibfield
  {journal} {\bibinfo  {journal} {Phys. Rev.E}\ }\textbf {\bibinfo {volume}
  {79}},\ \bibinfo {pages} {066109} (\bibinfo {year} {2009})}\BibitemShut
  {NoStop}%
\bibitem [{\citenamefont {Karmakar}\ \emph
  {et~al.}(2010{\natexlab{a}})\citenamefont {Karmakar}, \citenamefont
  {Lema{\^{i}}tre}, \citenamefont {Lerner},\ and\ \citenamefont
  {Procaccia}}]{10KLLP}%
  \BibitemOpen
  \bibfield  {author} {\bibinfo {author} {\bibfnamefont {S.}~\bibnamefont
  {Karmakar}}, \bibinfo {author} {\bibfnamefont {A.}~\bibnamefont
  {Lema{\^{i}}tre}}, \bibinfo {author} {\bibfnamefont {E.}~\bibnamefont
  {Lerner}}, \ and\ \bibinfo {author} {\bibfnamefont {I.}~\bibnamefont
  {Procaccia}},\ }\href {\doibase 10.1103/PhysRevLett.104.215502} {\bibfield
  {journal} {\bibinfo  {journal} {Phys. Rev. Lett.}\ }\textbf {\bibinfo
  {volume} {104}},\ \bibinfo {pages} {215502} (\bibinfo {year}
  {2010}{\natexlab{a}})}\BibitemShut {NoStop}%
\bibitem [{\citenamefont {Hentschel}\ \emph {et~al.}(2011)\citenamefont
  {Hentschel}, \citenamefont {Karmakar}, \citenamefont {Lerner},\ and\
  \citenamefont {Procaccia}}]{11HKLP}%
  \BibitemOpen
  \bibfield  {author} {\bibinfo {author} {\bibfnamefont {H.~G.~E.}\
  \bibnamefont {Hentschel}}, \bibinfo {author} {\bibfnamefont {S.}~\bibnamefont
  {Karmakar}}, \bibinfo {author} {\bibfnamefont {E.}~\bibnamefont {Lerner}}, \
  and\ \bibinfo {author} {\bibfnamefont {I.}~\bibnamefont {Procaccia}},\ }\href
  {\doibase 10.1103/PhysRevE.83.061101} {\bibfield  {journal} {\bibinfo
  {journal} {Phys. Rev.E}\ }\textbf {\bibinfo {volume} {83}},\ \bibinfo {pages}
  {061101} (\bibinfo {year} {2011})}\BibitemShut {NoStop}%
\bibitem [{\citenamefont {Dasgupta}\ \emph {et~al.}(2013)\citenamefont
  {Dasgupta}, \citenamefont {Hentschel}, \citenamefont {Procaccia},\ and\
  \citenamefont {Sen~Gupta}}]{13DHPS}%
  \BibitemOpen
  \bibfield  {author} {\bibinfo {author} {\bibfnamefont {R.}~\bibnamefont
  {Dasgupta}}, \bibinfo {author} {\bibfnamefont {H.~G.~E.}\ \bibnamefont
  {Hentschel}}, \bibinfo {author} {\bibfnamefont {I.}~\bibnamefont
  {Procaccia}}, \ and\ \bibinfo {author} {\bibfnamefont {B.}~\bibnamefont
  {Sen~Gupta}},\ }\href@noop {} {\bibfield  {journal} {\bibinfo  {journal} {EPL
  (Europhysics Letters)}\ }\textbf {\bibinfo {volume} {104}},\ \bibinfo {pages}
  {47003} (\bibinfo {year} {2013})}\BibitemShut {NoStop}%
\bibitem [{\citenamefont {Dubey}\ \emph {et~al.}(2015)\citenamefont {Dubey},
  \citenamefont {Hentschel}, \citenamefont {Jaiswal}, \citenamefont {Mondal},
  \citenamefont {Procaccia},\ and\ \citenamefont {Sen~Gupta}}]{15DHJMPS}%
  \BibitemOpen
  \bibfield  {author} {\bibinfo {author} {\bibfnamefont {A.~K.}\ \bibnamefont
  {Dubey}}, \bibinfo {author} {\bibfnamefont {H.~G.~E.}\ \bibnamefont
  {Hentschel}}, \bibinfo {author} {\bibfnamefont {P.~K.}\ \bibnamefont
  {Jaiswal}}, \bibinfo {author} {\bibfnamefont {C.}~\bibnamefont {Mondal}},
  \bibinfo {author} {\bibfnamefont {I.}~\bibnamefont {Procaccia}}, \ and\
  \bibinfo {author} {\bibfnamefont {B.}~\bibnamefont {Sen~Gupta}},\ }\href@noop
  {} {\bibfield  {journal} {\bibinfo  {journal} {EPL (Europhysics Letters)}\
  }\textbf {\bibinfo {volume} {112}},\ \bibinfo {pages} {17011} (\bibinfo
  {year} {2015})}\BibitemShut {NoStop}%
\bibitem [{\citenamefont {Hentschel}\ \emph {et~al.}(2016)\citenamefont
  {Hentschel}, \citenamefont {Procaccia},\ and\ \citenamefont
  {Sen~Gupta}}]{16HPS}%
  \BibitemOpen
  \bibfield  {author} {\bibinfo {author} {\bibfnamefont {H.~G.~E.}\
  \bibnamefont {Hentschel}}, \bibinfo {author} {\bibfnamefont {I.}~\bibnamefont
  {Procaccia}}, \ and\ \bibinfo {author} {\bibfnamefont {B.}~\bibnamefont
  {Sen~Gupta}},\ }\href@noop {} {\bibfield  {journal} {\bibinfo  {journal}
  {Phys. Rev.E}\ }\textbf {\bibinfo {volume} {93}},\ \bibinfo {pages} {033004}
  (\bibinfo {year} {2016})}\BibitemShut {NoStop}%
\bibitem [{\citenamefont {Gendelman}\ \emph {et~al.}(2016)\citenamefont
  {Gendelman}, \citenamefont {Lerner}, \citenamefont {Pollack}, \citenamefont
  {Procaccia}, \citenamefont {Rainone},\ and\ \citenamefont
  {Riechers}}]{16GLPPRR}%
  \BibitemOpen
  \bibfield  {author} {\bibinfo {author} {\bibfnamefont {O.}~\bibnamefont
  {Gendelman}}, \bibinfo {author} {\bibfnamefont {E.}~\bibnamefont {Lerner}},
  \bibinfo {author} {\bibfnamefont {Y.~G.}\ \bibnamefont {Pollack}}, \bibinfo
  {author} {\bibfnamefont {I.}~\bibnamefont {Procaccia}}, \bibinfo {author}
  {\bibfnamefont {C.}~\bibnamefont {Rainone}}, \ and\ \bibinfo {author}
  {\bibfnamefont {B.}~\bibnamefont {Riechers}},\ }\href {\doibase
  10.1103/PhysRevE.94.051001} {\bibfield  {journal} {\bibinfo  {journal} {Phys.
  Rev. E}\ }\textbf {\bibinfo {volume} {94}},\ \bibinfo {pages} {051001}
  (\bibinfo {year} {2016})}\BibitemShut {NoStop}%
\bibitem [{\citenamefont {Parisi}\ \emph {et~al.}(2018)\citenamefont {Parisi},
  \citenamefont {Pollack}, \citenamefont {Procaccia}, \citenamefont {Rainone},\
  and\ \citenamefont {Singh}}]{18PPPRS}%
  \BibitemOpen
  \bibfield  {author} {\bibinfo {author} {\bibfnamefont {G.}~\bibnamefont
  {Parisi}}, \bibinfo {author} {\bibfnamefont {Y.~G.}\ \bibnamefont {Pollack}},
  \bibinfo {author} {\bibfnamefont {I.}~\bibnamefont {Procaccia}}, \bibinfo
  {author} {\bibfnamefont {C.}~\bibnamefont {Rainone}}, \ and\ \bibinfo
  {author} {\bibfnamefont {M.}~\bibnamefont {Singh}},\ }\href {\doibase
  10.1103/PhysRevE.97.063003} {\bibfield  {journal} {\bibinfo  {journal} {Phys.
  Rev. E}\ }\textbf {\bibinfo {volume} {97}},\ \bibinfo {pages} {063003}
  (\bibinfo {year} {2018})}\BibitemShut {NoStop}%
\bibitem [{\citenamefont {Mezard}\ and\ \citenamefont {Parisi}(2009)}]{09MG}%
  \BibitemOpen
  \bibfield  {author} {\bibinfo {author} {\bibfnamefont {M.}~\bibnamefont
  {Mezard}}\ and\ \bibinfo {author} {\bibfnamefont {G.}~\bibnamefont
  {Parisi}},\ }\href@noop {} {\enquote {\bibinfo {title} {Glasses and
  replicas},}\ } (\bibinfo {year} {2009}),\ \Eprint
  {http://arxiv.org/abs/0910.2838v1} {arXiv:0910.2838v1} \BibitemShut {NoStop}%
\bibitem [{\citenamefont {Dubey}\ \emph {et~al.}(2016)\citenamefont {Dubey},
  \citenamefont {Procaccia}, \citenamefont {Shor},\ and\ \citenamefont
  {Singh}}]{16DPSS}%
  \BibitemOpen
  \bibfield  {author} {\bibinfo {author} {\bibfnamefont {A.~K.}\ \bibnamefont
  {Dubey}}, \bibinfo {author} {\bibfnamefont {I.}~\bibnamefont {Procaccia}},
  \bibinfo {author} {\bibfnamefont {C.~A.}\ \bibnamefont {Shor}}, \ and\
  \bibinfo {author} {\bibfnamefont {M.}~\bibnamefont {Singh}},\ }\href
  {\doibase 10.1103/PhysRevLett.116.085502} {\bibfield  {journal} {\bibinfo
  {journal} {Phys. Rev. Lett.}\ }\textbf {\bibinfo {volume} {116}},\ \bibinfo
  {pages} {085502} (\bibinfo {year} {2016})}\BibitemShut {NoStop}%
\bibitem [{\citenamefont {Kob}\ and\ \citenamefont {Andersen}(1995)}]{95KA}%
  \BibitemOpen
  \bibfield  {author} {\bibinfo {author} {\bibfnamefont {W.}~\bibnamefont
  {Kob}}\ and\ \bibinfo {author} {\bibfnamefont {H.~C.}\ \bibnamefont
  {Andersen}},\ }\href {\doibase 10.1103/PhysRevE.52.4134} {\bibfield
  {journal} {\bibinfo  {journal} {Phys. Rev.E}\ }\textbf {\bibinfo {volume}
  {52}},\ \bibinfo {pages} {4134} (\bibinfo {year} {1995})},\ \Eprint
  {http://arxiv.org/abs/9505118} {9505118} \BibitemShut {NoStop}%
\bibitem [{\citenamefont {Br{\"{u}}ning}\ \emph {et~al.}(2009)\citenamefont
  {Br{\"{u}}ning}, \citenamefont {St-Onge}, \citenamefont {Patterson},\ and\
  \citenamefont {Kob}}]{09BSPK}%
  \BibitemOpen
  \bibfield  {author} {\bibinfo {author} {\bibfnamefont {R.}~\bibnamefont
  {Br{\"{u}}ning}}, \bibinfo {author} {\bibfnamefont {D.~A.}\ \bibnamefont
  {St-Onge}}, \bibinfo {author} {\bibfnamefont {S.}~\bibnamefont {Patterson}},
  \ and\ \bibinfo {author} {\bibfnamefont {W.}~\bibnamefont {Kob}},\ }\href
  {\doibase 10.1088/0953-8984/21/3/035117} {\bibfield  {journal} {\bibinfo
  {journal} {J. Phys.: Cond. Matt.}\ }\textbf {\bibinfo {volume} {21}},\
  \bibinfo {pages} {035117} (\bibinfo {year} {2009})}\BibitemShut {NoStop}%
\bibitem [{\citenamefont {Karmakar}\ \emph
  {et~al.}(2010{\natexlab{b}})\citenamefont {Karmakar}, \citenamefont {Lerner},
  \citenamefont {Procaccia},\ and\ \citenamefont {Zylberg}}]{10KLPZ}%
  \BibitemOpen
  \bibfield  {author} {\bibinfo {author} {\bibfnamefont {S.}~\bibnamefont
  {Karmakar}}, \bibinfo {author} {\bibfnamefont {E.}~\bibnamefont {Lerner}},
  \bibinfo {author} {\bibfnamefont {I.}~\bibnamefont {Procaccia}}, \ and\
  \bibinfo {author} {\bibfnamefont {J.}~\bibnamefont {Zylberg}},\ }\href
  {\doibase 10.1103/PhysRevE.82.031301} {\bibfield  {journal} {\bibinfo
  {journal} {Phys. Rev.E}\ }\textbf {\bibinfo {volume} {82}},\ \bibinfo {pages}
  {031301} (\bibinfo {year} {2010}{\natexlab{b}})},\ \Eprint
  {http://arxiv.org/abs/1006.3737} {arXiv:1006.3737} \BibitemShut {NoStop}%
\bibitem [{\citenamefont {Lutsko}(1989)}]{89Lutsko}%
  \BibitemOpen
  \bibfield  {author} {\bibinfo {author} {\bibfnamefont {J.~F.}\ \bibnamefont
  {Lutsko}},\ }\href {\doibase 10.1063/1.342716} {\bibfield  {journal}
  {\bibinfo  {journal} {Journal of Applied Physics}\ }\textbf {\bibinfo
  {volume} {65}},\ \bibinfo {pages} {2991} (\bibinfo {year}
  {1989})}\BibitemShut {NoStop}%
\bibitem [{\citenamefont {Barrat}(2006)}]{barrat2006microscopic}%
  \BibitemOpen
  \bibfield  {author} {\bibinfo {author} {\bibfnamefont {J.-L.}\ \bibnamefont
  {Barrat}},\ }in\ \href@noop {} {\emph {\bibinfo {booktitle} {Computer
  Simulations in Condensed Matter Systems: From Materials to Chemical Biology
  Volume 2}}}\ (\bibinfo  {publisher} {Springer},\ \bibinfo {year} {2006})\
  pp.\ \bibinfo {pages} {287--307}\BibitemShut {NoStop}%
\bibitem [{\citenamefont {Yoshino}(2012)}]{12Yos}%
  \BibitemOpen
  \bibfield  {author} {\bibinfo {author} {\bibfnamefont {H.}~\bibnamefont
  {Yoshino}},\ }\href {\doibase 10.1063/1.4722343} {\bibfield  {journal}
  {\bibinfo  {journal} {J. Chem. Phys.}\ }\textbf {\bibinfo {volume} {136}},\
  \bibinfo {pages} {214108} (\bibinfo {year} {2012})}\BibitemShut {NoStop}%
\bibitem [{\citenamefont {Henkes}\ \emph {et~al.}(2012)\citenamefont {Henkes},
  \citenamefont {Brito},\ and\ \citenamefont
  {Dauchot}}]{dauchot2012pedagogical}%
  \BibitemOpen
  \bibfield  {author} {\bibinfo {author} {\bibfnamefont {S.}~\bibnamefont
  {Henkes}}, \bibinfo {author} {\bibfnamefont {C.}~\bibnamefont {Brito}}, \
  and\ \bibinfo {author} {\bibfnamefont {O.}~\bibnamefont {Dauchot}},\
  }\href@noop {} {\bibfield  {journal} {\bibinfo  {journal} {Soft Matter}\
  }\textbf {\bibinfo {volume} {8}},\ \bibinfo {pages} {6092} (\bibinfo {year}
  {2012})}\BibitemShut {NoStop}%
\bibitem [{\citenamefont {Stoessel}\ and\ \citenamefont
  {Wolynes}(1984)}]{84SW}%
  \BibitemOpen
  \bibfield  {author} {\bibinfo {author} {\bibfnamefont {J.~P.}\ \bibnamefont
  {Stoessel}}\ and\ \bibinfo {author} {\bibfnamefont {P.~G.}\ \bibnamefont
  {Wolynes}},\ }\href {\doibase 10.1063/1.447235} {\bibfield  {journal}
  {\bibinfo  {journal} {J. Chem. Phys.}\ }\textbf {\bibinfo {volume} {80}},\
  \bibinfo {pages} {4502} (\bibinfo {year} {1984})}\BibitemShut {NoStop}%
\bibitem [{\citenamefont {Hall}\ and\ \citenamefont {Wolynes}(2008)}]{08RW}%
  \BibitemOpen
  \bibfield  {author} {\bibinfo {author} {\bibfnamefont {R.~W.}\ \bibnamefont
  {Hall}}\ and\ \bibinfo {author} {\bibfnamefont {P.~G.}\ \bibnamefont
  {Wolynes}},\ }\href@noop {} {\bibfield  {journal} {\bibinfo  {journal} {J.
  Phys. Chem. B}\ }\textbf {\bibinfo {volume} {112}},\ \bibinfo {pages} {301}
  (\bibinfo {year} {2008})}\BibitemShut {NoStop}%
\end{thebibliography}%

\end{document}